\documentstyle[11pt,paspconf,epsf]{article}
\begin{document}

\title{Deep ISOCAM Observations of Abell 2218}

\author{B. Altieri, L.Metcalfe, A. Biviano\altaffilmark{1}, 
        K. Leech, K. Okumura \& B. Schulz }
\affil{ISO Science Operations Centre, Astrophysics Division, Space Science
       Dept. of ESA, Villafranca, P.O. Box 50727, Madrid, Spain}

\author{B. Mc Breen, M. Delaney}
\affil{Department of Experimental Physics, University College, Dublin,
       Ireland}

\altaffiltext{1}{Osservatorio Astronomico di Trieste, Italy} 

\begin{abstract}
We report on the mid-infrared imaging at 5, 7, 10 \& 15 $\mu$m of the
galaxy cluster Abell 2218 obtained with the ISOCAM instrument onboard ESA's
Infrared Space Observatory (ISO), as part of an on-going program to image
gravitational arcs and arclets in distant clusters. Several cluster galaxies
as well as field galaxies are detected. We discuss their mid-IR flux 
properties.

\end{abstract}


\keywords{gravitational lensing, galaxy clusters}

\section{Introduction}
Abell 2218 is a very 
rich galaxy cluster (richness class 4, according to Abell et al. 1989),
characterized by a large velocity dispersion of the galaxy population 
($\sigma_v=1370$~km/s, Le Borgne et al. 1992), and a high X-ray temperature
and luminosity ($T_x=7.2$~keV, see e.g. Markevitch 1997, 
$L_x = 2.9 \times 10^{44}$ 
erg/s\footnote{In this paper we adopt $H_0=75$~km/s/Mpc and $q_0=1/2$}
in the energy band 0.5--4.4 keV, see e.g. Kneib et al. 1996). 

These properties, coupled to a relatively small distance 
($<z>=0.175$, Le Borgne et al. 1992), made this cluster an attractive target for
studies of the Sunyaev-Zeldovich effect (Birkinshaw \& Hughes 1994), and
one of the closest clusters where gravitational arcs are detected. The amazing
concentration of gravitational arcs and arclets has stimulated a huge 
observational effort first to get sub-arcsec imaging (from the ground, 
see, e.g. Kneib et al. 1995, and from space with HST, Kneib et al. 1996), 
and then to get spectra of the arcs (Ebbels et al. 1997). 

The optical and near-IR observations have allowed a very detailed modelling 
of the mass distribution within the cluster (Kneib et al. 1996), confirmed
later to a great level of accuracy, (Ebbels et al. 1997). 
It was found that the cluster mass
distribution is bi-modal, with the main concentration centred on the cD
galaxy. The X-ray surface brightness, as obtained via ROSAT observations,
does not trace the gravitational potential as derived from the lensing
analysis, a possible indication that the X-ray emitting gas is far from
hydrostatic equilibrium (Kneib et al. 1996). This would also explain the
discrepancy in the X-ray and lensing mass estimates (Markevitch 1997).

In this paper we report on the mid-IR observations of Abell 2218 done with 
ISOCAM on-board ESA's {\em Infrared Space Observatory} (ISO) satellite. 
The high sensitivity of ISOCAM allows us to determine the photometric
properties of lensed and cluster galaxies in the mid-IR band, thus 
widening the wavelength coverage of their spectra. Mid-IR observations
are critical in understanding the intrinsic nature of these distant galaxies,
since a starburst is known to emit a large fraction of its energy in the mid-IR,
due to dust grain re-processing of the emitted radiation.

\section{Observations and data reductions}

Abell 2218 was observed on March 3, 1996 and again on February 28, 1997
as part of an ISO program 
of observations of lensing galaxy clusters with arcs
and arclets: Abell 370, Abell 2218, Cl2244-02 and MS2137-23. An important
criterion for target selection was to choose the clusters with the
brightest arcs in the optical, spatially extended and the farthest from
the lens, due to the limited spatial resolution of ISO.\\
The observations were performed by rastering in microscanning mode yielding
to very deep ISOCAM images of the galaxy cluster. The pixel field-of-view was
3$^{\prime\prime}$ and the microscanning steps were 7$^{\prime\prime}$ 
in both directions of the 5x5 raster.
The cluster was imaged in the 4 filter bands, LW1 [4-5$\mu$m], 
LW2 [5-8.5$\mu$m], LW7 [8.5-11$\mu$m],
and LW3 [12-18$\mu$m],
at 5 or 10 sec integration
times for a total observing time of 5.5 hours, covering a field of more
than 2x2 arcminutes.
The data reduction was done partly within the IDL based (ISO)CAM Interactive
Analysis (CIA) package (Ott et al., 1997) and partly through the usage of 
C++ based Multi-resolution Median Transform ({\em wavelet}) techniques
(Starck J.-L. et al, 1997).
Cross-correlation of the two methods allows us to
increase the reliability of source detection.

\section{Results and discussion}

The LW1, LW2 and LW3 maps are presented in figures 1 \& 2a, 
as overlays on top of the HST image (courtesy J.-P. Kneib) of Abell 2218.

\begin{figure}
\plottwo{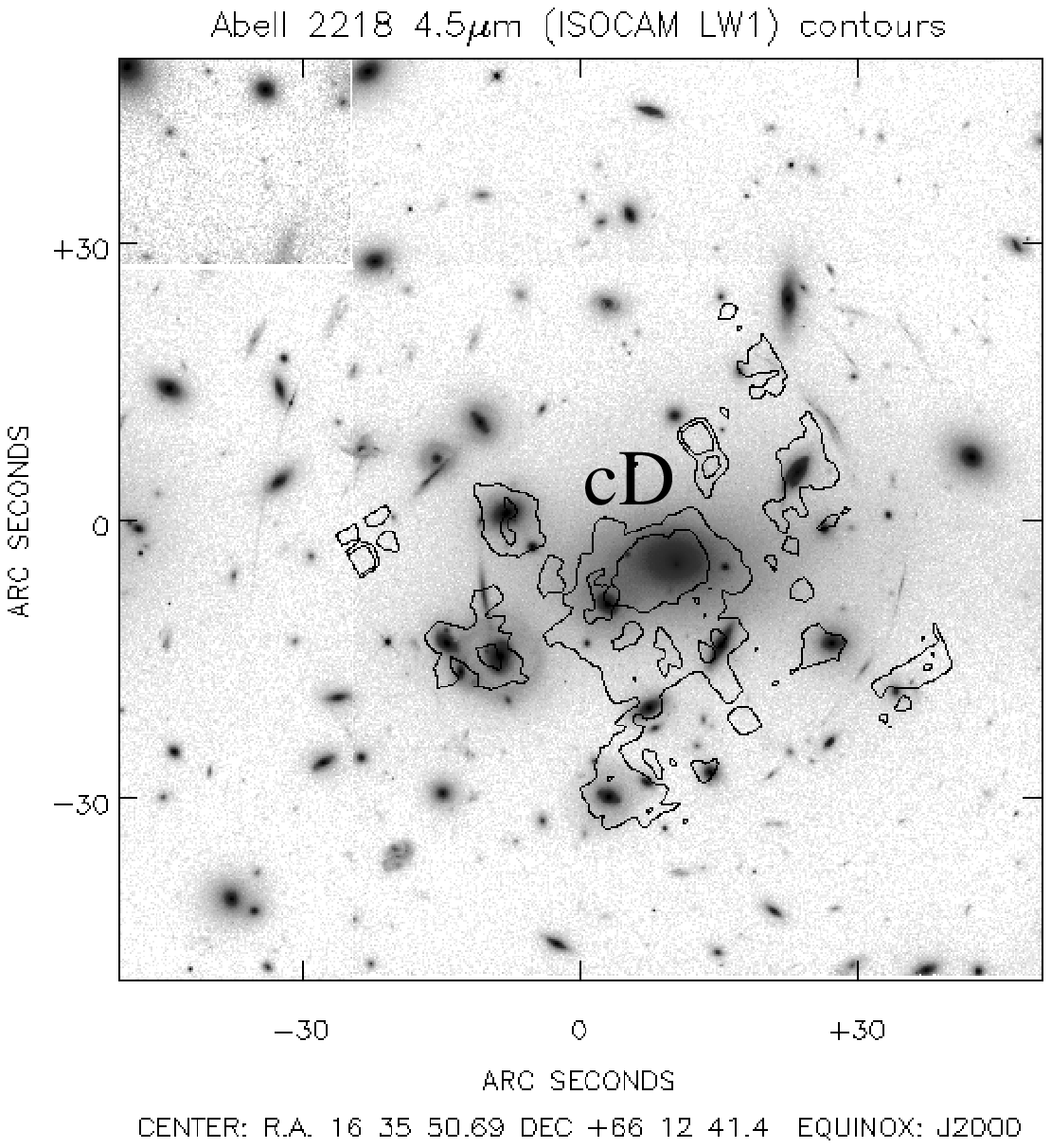}{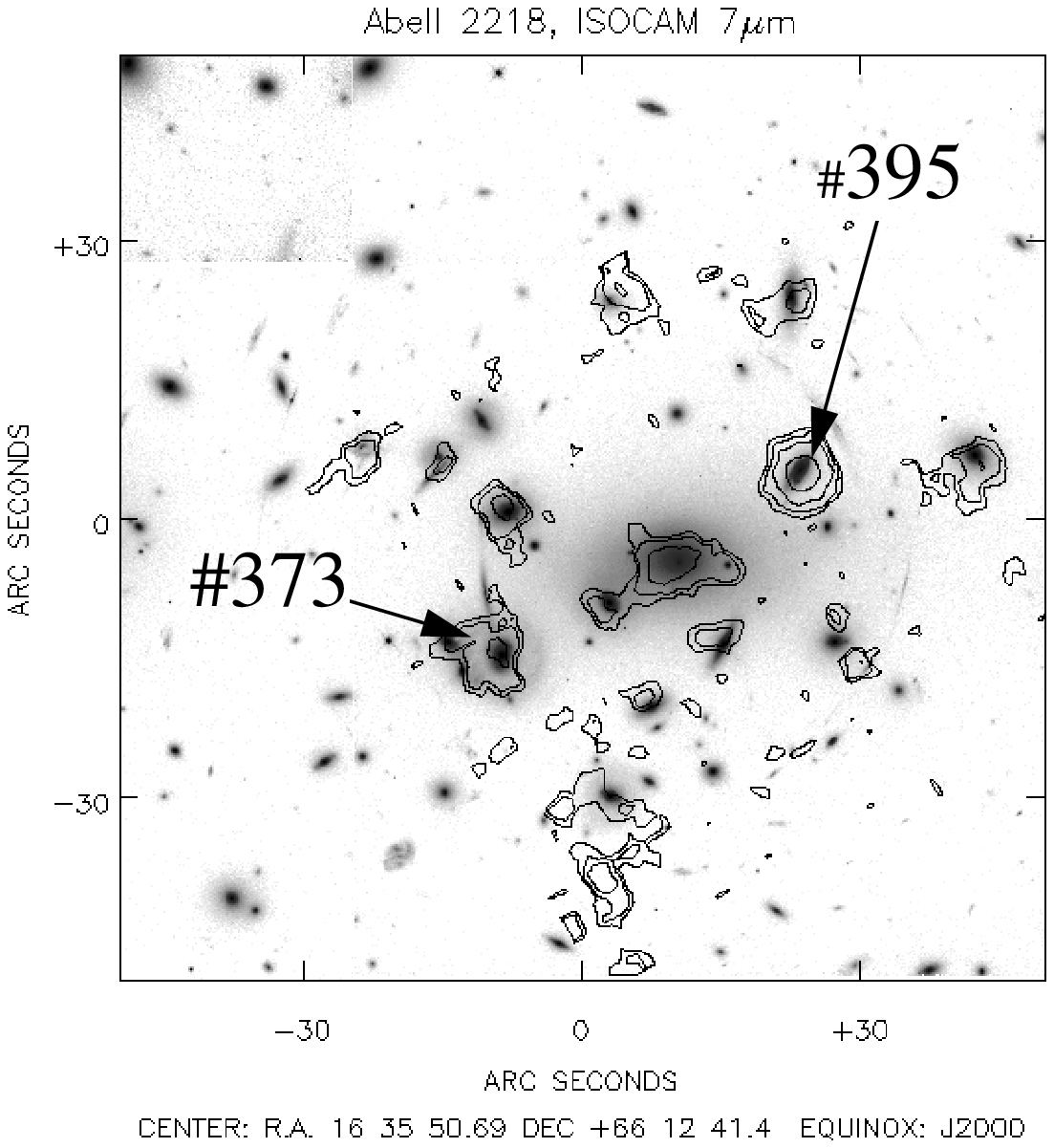}
\caption{Abell 2218 4.5$\mu$m (a) \& 7$\mu$m (b) contour maps}
\label{fig-1}
\end{figure}

In the cluster core, no arclet is detected at a significant level at
any wavelength in the mid-IR. This non-detection of arcs or arclets
contrasts with our recent mid-IR imaging of Abell 370 (Metcalfe et al. 1997), 
where the A0 giant arc is clearly detected as the main feature 
and emitter in the cluster core at 15$\mu$m. But this giant arc
is already a prominent feature in the optical, (first
giant arc discovered), whereas arcs and arclets are fainter in Abell 2218. \\
The cD galaxy which is apparently centred on the cluster potential is clearly
detected in the LW1 \& LW2 filter bands. 
At 4.5$\mu$m (rest-frame wavelength 
3.8 $\mu$m) the cD emission is extended,
covering part of the optical halo; 
at 7$\mu$m the emission is much more confined
to the centre, but with an extension along the optical major axis, probably
contaminated by the neighbouring merging dwarf galaxies. At 10$\mu$m and 
15$\mu$m it lies just above the detection limit of ISOCAM and no statement 
can be made on its
extension. The mid-IR  spectral energy distribution up to 15$\mu$m seems to 
follow a simple Rayleigh-Jeans tail of the cold stellar component as found in 
optically-selected 
normal early-type galaxies (E, S0, S0a) in the Virgo cluster 
(Boselli et al. 1998).
But due to the contamination by close-by galaxies it is very difficult to
estimate the 4.5$\mu$m flux of the cD. \\
The brightest cluster member galaxies are also detected at 4.5 $\mu$m and 
7 $\mu$m.
At the redshift of the cluster the 4.5 $\mu$m emission is mostly 
coming again from cold stellar photospheres.
The 7 $\mu$m (LW2 band) (rest-frame 5.7 $\mu$m)
emission includes a small fraction of PAH emission but comes mainly again 
from the cold stellar photospheres. 
At 10 \& 15 microns most of cluster members have vanished  
(ie. are below or very close to our detection limit), as expected for normal 
early-type galaxies (ellipticals) where there is only the 
Rayleigh-Jeans tail mid-IR emission from cold stellar photospheres, 
in dust \& gas poor cluster core ellipticals.\\

Still one cluster member galaxy \#373 (numbering system from Le Borgne 
et al. 1992 in the following) is detected at 15$\mu$m, but also
at 4.5 $\mu$m and 7 $\mu$m, it has a 'cart-wheel' like aspect, from the ring
surrounding it and is probably a big face-on spiral galaxy. The 4.5$\mu$m and 
7$\mu$m emission comes mostly from the nucleus whereas the 15$\mu$m 
originates apparently from the 'ring' structure.
It could be PAH and small hot grain emission enhanced by star formation
in tidal shocks in the 'ring'.

\begin{figure}
\plottwo{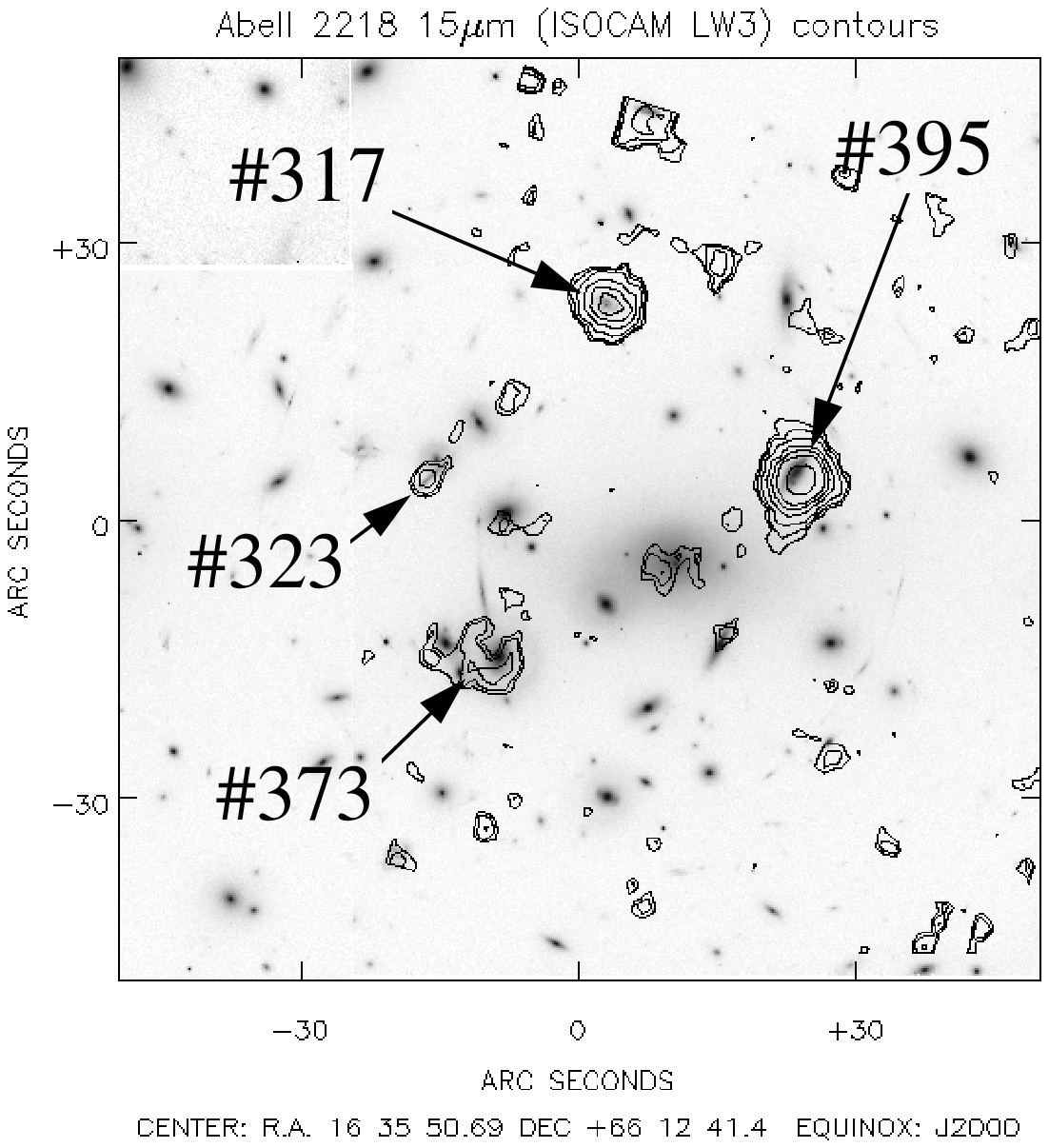}{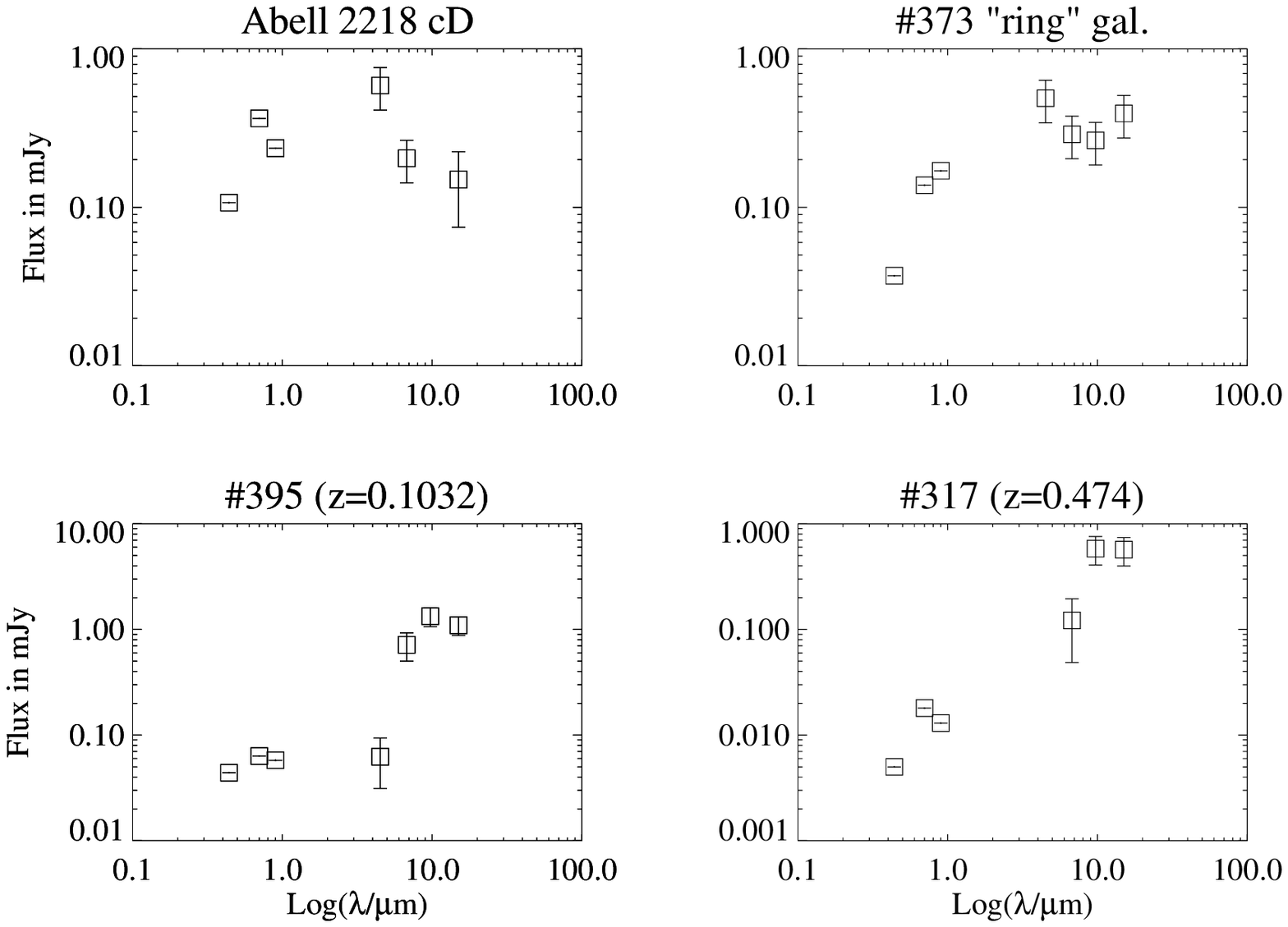}
\caption{Abell 2218 15$\mu$m contour maps (a) and spectral energy distributions
of 4 objects}
\label{fig-2}
\end{figure}

Object \#323 is detected at 15 $\mu$m. 
This object was suspected to be an arclet 
candidate in the first deep multi-filter ground-based imaging 
(Pello et al. 1992), since it appeared as a fine unresolved extended 
structure, together with another 32 arclets in total. 
It is classified as one of the 235 arclet candidates in the
first HST observation of Abell 2218 (Kneib et al. 1996).
However, as noted by Kneib et al., it was also suspected not to be a strongly 
lensed object when comparison is made with shear orientation. 
This was again confirmed by Ebbels et al. (1997), object \#323 being very 
elongated but at an angle of 45 degrees from the shear direction. 
Identification of several
absorption features in its spectrum reveal it to be a cluster member at a 
redshift of z=0.179, a spiral seen edge-on.

To our surprise the brightest source by far from 7$\mu$m to 15$\mu$m is object
\#395, an apparently insignificant z=0.1032 (Le borgne 1992) foreground Sb 
galaxy, showing an ultraviolet excess and a strong H$\alpha$ emission line, 
both indicating strong star formation of massive stars. 
Most of this mid-IR emission could then originate from the reprocessing 
by the small hot dust heated by a strong UV radiation field.

The second strongest emitter at long wavelength (from 10$\mu$m to 15$\mu$m),
but much less at shorter wavelengths is object \#317, a lensed galaxy at
z=0.474 (Ebbels et al. 1997), a rather red object from the visible to
the mid-IR. First indications show that this type of lensed object is
common on our larger field imaging towards lensing clusters (Metcalfe et al.
1998 in preparation)

The spectral energy distributions from the visible (B,R,I) 
to the mid-IR of the 4 objects discussed above are shown in fig. 2b

\end{document}